 \documentclass{emulateapj}

\newcommand{\myemail}{enomoto@icrr.u-tokyo.ac.jp}

\slugcomment{to appear in ApJ.}

\shorttitle{A Search for TeV Gamma-rays from the Vela Pulsar Region}
\shortauthors{Enomoto et al.}

\begin{document}

\title{A Search for sub-TeV Gamma-rays from the 
Vela Pulsar Region with CANGAROO-III}

\author{
R.~Enomoto\altaffilmark{1}, 
K.~Tsuchiya\altaffilmark{1}, 
Y.~Adachi\altaffilmark{1}, 
S.~Kabuki\altaffilmark{2},  
P.G.~Edwards\altaffilmark{3}, 
A.~Asahara\altaffilmark{2}, 
G.V.~Bicknell\altaffilmark{4}, 
R.W.~Clay\altaffilmark{5}, 
Y.~Doi\altaffilmark{6},   
S.~Gunji\altaffilmark{6}, 
S.~Hara\altaffilmark{1}, 
T.~Hara\altaffilmark{7}, 
T.~Hattori\altaffilmark{8}, 
Sei.~Hayashi\altaffilmark{9},  
Y.~Higashi\altaffilmark{2},
R.~Inoue\altaffilmark{8}, 
C.~Itoh\altaffilmark{10}, 
F.~Kajino\altaffilmark{9}, 
H.~Katagiri\altaffilmark{2}, 
A.~Kawachi\altaffilmark{8}, 
S.~Kawasaki\altaffilmark{1},
T.~Kifune\altaffilmark{11},
R.~Kiuchi\altaffilmark{1}, 
K.~Konno\altaffilmark{6},
L.T.~Ksenofontov\altaffilmark{1}, 
H.~Kubo\altaffilmark{2}, 
J.~Kushida\altaffilmark{8}, 
Y.~Matsubara\altaffilmark{12}, 
Y.~Mizumoto\altaffilmark{13},
M.~Mori\altaffilmark{1}, 
H.~Muraishi\altaffilmark{14},
Y.~Muraki\altaffilmark{12},
T.~Naito\altaffilmark{7},
T.~Nakamori\altaffilmark{2}, 
D.~Nishida\altaffilmark{2}, 
K.~Nishijima\altaffilmark{8}, 
M.~Ohishi\altaffilmark{1}, 
J.R.~Patterson\altaffilmark{5}, 
R.J.~Protheroe\altaffilmark{5}, 
Y.~Sakamoto\altaffilmark{6},   
M.~Sato\altaffilmark{6}, 
S.~Suzuki\altaffilmark{15},
T.~Suzuki\altaffilmark{15},
D.L.~Swaby\altaffilmark{5}, 
T.~Tanimori\altaffilmark{2}, 
H.~Tanimura\altaffilmark{2}, 
G.J.~Thornton\altaffilmark{5}, 
S.~Watanabe\altaffilmark{2},  
T.~Yamaoka\altaffilmark{6},  
M.~Yamazaki\altaffilmark{9},  
S.~Yanagita\altaffilmark{15}, 
T.~Yoshida\altaffilmark{15},
T.~Yoshikoshi\altaffilmark{1},
M.~Yuasa\altaffilmark{1},
Y.~Yukawa\altaffilmark{1}}

\altaffiltext{1}
{ICRR, Univ.\ of Tokyo, Kashiwa, Chiba 277-8582, Japan; \myemail}
\altaffiltext{2}
{Department of Physics, Graduate School of Science, Kyoto University, 
Sakyo-ku, Kyoto 606-8502, Japan}
\altaffiltext{3}
{Institute of Space and Astronautical Science, 
Japan Aerospace Exploration Agency, Sagamihara, 
Kanagawa 229-8510, Japan} 
\altaffiltext{4}
{Research School of Astronomy and Astrophysics, 
 Australian National University, ACT 2611, Australia}
\altaffiltext{5}
{Department of Physics and Mathematical Physics, University of
Adelaide, SA 5005, Australia}
\altaffiltext{6}
{Department of Physics, Yamagata University, Yamagata, 
Yamagata 990-8560, Japan}
\altaffiltext{7}
{Faculty of Management Information, Yamanashi Gakuin University, 
Kofu, Yamanashi 400-8575, Japan}
\altaffiltext{8}
{Department of Physics, Tokai University, Hiratsuka, 
Kanagawa 259-1292, Japan}
\altaffiltext{9}
{Department of Physics, Konan University, Kobe, Hyogo 658-8501, Japan}
\altaffiltext{10}
{Department of Medical Imaging, National Institute of Radiological Sciences, Chiba, 
Chiba 263-8555, Japan} 
\altaffiltext{11}
{Faculty of Engineering, Shinshu University, Nagano, Nagano 480-8553, Japan}
\altaffiltext{12}
{Solar-Terrestrial Environment Laboratory,  Nagoya University, 
Nagoya, Aichi 464-8602, Japan}
\altaffiltext{13}
{National Astronomical Observatory of Japan, Mitaka, Tokyo 181-8588, Japan}
\altaffiltext{14}
{School of Allied Health Sciences, Kitasato University, Sagamihara, 
Kanagawa 228-8555, Japan}
\altaffiltext{15}
{Faculty of Science, Ibaraki University, Mito, Ibaraki 310-8512, Japan}

\begin{abstract}

We made stereoscopic observations of the Vela Pulsar region
with two of the  10\,m diameter CANGAROO-III imaging
atmospheric Cherenkov telescopes in January and February, 2004,
in a search for sub-TeV gamma-rays from the pulsar and surrounding
regions. We describe the observations, provide a detailed account
of the calibration methods, 
and introduce the 
improved and bias-free analysis techniques employed for CANGAROO-III data.
No evidence of gamma-ray emission is found from either the pulsar
position or the previously reported position offset by 0.13~degree, 
and the resulting upper limits are
a factor of five less than the previously
reported flux from observations with the CANGAROO-I 3.8\,m telescope.
Following the recent report by the H.E.S.S.\ group of
TeV gamma-ray emission from 
the Pulsar Wind Nebula,
which is $\sim$0.5~degree south of the pulsar position,
we examined this region and found supporting evidence for emission
extended over $\sim$0.6~degree.

\end{abstract}

\keywords{gamma rays: observations --- pulsars: individual (Vela) --- 
methods: data analysis}

\section{Introduction}

The Vela Pulsar is the brightest object in the sky at 100\,MeV energies 
\citep{3EGcatalog}, with emission extending beyond 10\,GeV 
\citep{thompson}. 
The emission at these energies is totally pulsed \citep{kanbach}.
The pulsar 
undergoes large glitches \citep[e.g.,][]{dodson}, and
has one of the highest values of
$\dot{E}$/d$^2$ of all pulsars, 
making it a prime target for southern
hemisphere searches for TeV gamma-rays \citep[e.g.,][]{nel}.
Early searches for TeV gamma-ray emission relied on the detection
of a pulsed signal, and produced a number of
suggestive but inconsistent results \citep[see][for a review]{pge}.

The first search sensitive to both pulsed and steady emission
and made using the proven imaging atmospheric Cherenkov technique
was undertaken with the CANGAROO-I 3.8\,m telescope.
A total of 119 hours of usable on-source data taken between 1993 and 1995
yielded a 5.8$\sigma$
excess, corresponding to a flux of 
(2.9$\pm$0.5$\pm$0.4) $\times$ 10$^{-12}$ cm$^{-2}$s$^{-1}$ above 
2.5$\pm$1.0\,TeV,
offset from the
Vela Pulsar by 0.13~degrees and with no significant modulation
at the pulsar period  \citep{yoshikoshi}.
A further 29\,hours data from 
CANGAROO-I observations in 1997 showed a 4.1$\sigma$
excess at a position consistent with peak from the
earlier observations \citep{yoshikoshi,yoshikoshi2}.
Observations by the Durham group in 1996 produced 3$\sigma$ upper limits above
300\,GeV of $5\times 10^{-11}$ and $1.3\times 10^{-11}$\,cm$^{-2}$s$^{-1}$
for steady and pulsed emission, respectively, at the
pulsar position \citep{chadwick}.
No significant excess was found in a 2$^\circ \times$2$^\circ$ area
centered at the pulsar position.

Based on independent Monte-Carlo simulations of the CANGAROO-I
telescope \citep{dazeley1}, \cite{dazeley2} searched for a
TeV gamma-ray signal in the 29\,hour 1997 dataset,
but found no evidence of emission. \cite{dazeley2} noted that increasing the
value of the {\it length} cut toward the value used in
the earlier analysis did result in a somewhat increased excess, though
not to a significant level, and acknowledged that the
simulations had not included several known characteristics of the
CANGAROO-I telescope hardware. An analysis of data from the 
Crab Nebula taken between December 1996 and January 1998 
with two different sets of cuts also
failed to yield a significant excess \citep{dazeley2}, although
the resulting upper limit was higher than the established flux
of the Crab Nebula. The TeV flux from the Crab Nebula from
earlier CANGAROO-I \citep{tanimori} observations is consistent with 
the most recent measurements \citep{hegra2004}.

The first preliminary results from observations with the H.E.S.S.\ 
stereoscopic imaging Cherenkov telescopes showed no evidence
of emission from the Vela Pulsar nor from previous CANGAROO-I position 
\citep{gamma2004}. 
However, very recently the H.E.S.S.\ group reported the detection
of broad TeV emission 
$\sim$0.5~degree south from the pulsar position
extended over a region $\lesssim$ 0.6~degree \citep{icrc2005}.
The preliminary non-detection of emission from the pulsar was
confirmed, yet the extended emission from the region of the Vela
pulsar wind nebula was at the level of 0.5~crab
\citep{icrc2005}.
Here we report the results from our observations of the Vela Pulsar region
made in early of 2004
with the new CANGAROO-III stereoscopic imaging atmospheric system.

The previous CANGAROO-I analyses used slightly different cut values, 
especially in the image analyses,
due in part to differences in source declination and sky brightness.
Analysis using three or four image parameters requires
selection of six to eight cut parameters
(i.e., both upper- and lower-cut values for each parameter).
While the cuts were guided by simulations,
which endeavored to incorporate the characterstics of
all telescope hardware and electronics \citep[e.g.,][]{dazeley1},
any fine tuning introduces a number of degrees of freedom into the analysis,
which can be difficult to account for a posteriori.
The number of parameters can be reduced to one
with the use of mathematical methods such as
Likelihood \citep{enomoto_nature}
or the Fisher Discriminant \citep{fisher}.
In this report, we emphasize the application of such methods
in order to remove any potentially biased operation from the analysis
\citep[see. e.g.,][]{weekes}.
Previous CANGAROO-II analyses have been carried out
using the Likelihood ratio, a Likelihood analysis with only 
one image cut \citep{enomoto_nature}. 
The Fisher Discriminant analysis introduced here
now removes all cut uncertainties from the image analysis.

In \S2 we describe the CANGAROO-III system, and in \S3 we present the
various calibration checks made to confirm its performance, 
introducing unbiased analysis methods based on
Likelihood \citep{enomoto_nature} and the Fisher Discriminant
\citep{fisher}. In \S\S4 and 5  the results of the CANGAROO-III
observations of the Vela Pulsar and the surrounding field are presented,
with conclusions given in \S6.

\section{The CANGAROO-III Stereoscopic System}

The imaging atmospheric Cherenkov technique (IACT) was
pioneered with the Whipple group's detection of TeV gamma-rays 
from the Crab Nebula \citep{whipple}.
This technique enabled TeV gamma-rays to be distinguished 
from the huge background of cosmic rays with the use of the
``Hillas moments" of the Cherenkov images \citep{hillas}.
Stereoscopic observations were successfully demonstrated by the
HEGRA group to significantly improve the rejection of the
cosmic ray background \citep{hegra}.
More recently, the H.E.S.S.\ group has reported the detection of 
a number of faint gamma-ray sources 
with angular resolutions of only a few-arc minutes \citep{HESS_science}.

The CANGAROO-III stereoscopic system consists of four imaging atmospheric
imaging telescopes located near Woomera, South Australia (31$^\circ$06'S,
136$^\circ$47'E, 160 m a.s.l.).
Each telescope has a 10\,m diameter reflector
consisting of 114 spherical mirror segments, each of 80\,cm
diameter and with an average radius of curvature of 16.4\,m.
The segments are made of Fiber Reinforced Plastic (FRP)
\citep{kawachi} and are aligned on a parabolic
frame  with a focal length of 8\,m.
The total light collecting area is 57\,m$^2$.
The four telescopes are located on the corners of a diamond 
with spacings of approximately 100\,m \citep{enomoto_app}.
The first telescope, designated T1, is the CANGAROO-II
telescope, and data from this telescope was not used in the analysis 
presented here 
as its smaller field of view (FOV) was less well suited for these 
stereoscopic observations.
The second and third CANGAROO-III telescopes, T2 and T3, were 
used for the observations described here, with the fourth telescope (T4) 
becoming operational after these observations were completed.
The camera systems for T2, T3, and T4 are identical and are
described in detail in \cite{kabuki}.
The pixel timing resolution is 1\,ns at 20 photoelectron (p.e.) input.
The noise level is 0.1\,p.e.
The relative gains of pixels are adjusted to within 10\% at the
hardware level
by adjusting the high voltages applied to photomultipliers.
The offline calibration are carried out using LED flasher
data which were taken once per day. The gain uniformity
of pixels is believed to be less than 5\%.

\begin{figure}
\epsscale{1.0}
\plotone{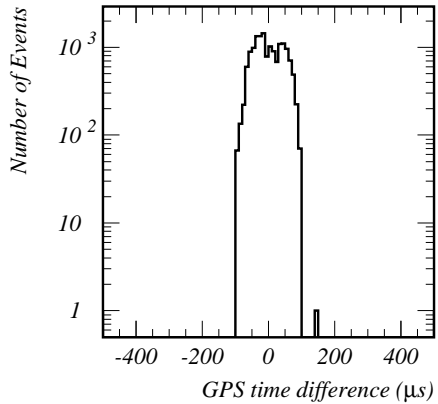}
\caption{
GPS time differences in units of $\mu$s for all events.
}
\label{f_gps}
\end{figure}

The observations were made using the so-called ``wobble mode"
in which the pointing position of each telescope was
alternated between $\pm$0.5~degree in declination from the 
center of the target every 20 minutes \citep{wobble}.
Two telescopes, T2 and T3, were operational at this time, with
T3 located 100\,m to the south-south-east of T2.
The data from each telescope were recorded with the trigger condition
requiring that more than four pixels exceeded 7.6~p.e.
The GPS time stamp was also recorded for each event.
In the offline analysis, we combine these datasets by comparing
the GPS timing.
The distribution of timing differences is shown in Fig.~\ref{f_gps}.
The broadness of the distribution was due to the time resolution 
of the GPS clocks.
We required an off-line coincidence between the two telescopes' 
trigger times within $\pm$100\,$\mu$s.
Events outside this time window were considered to be accidental
coincidences.
The trigger rates of the individual telescopes were at most 80\,Hz 
and this was
reduced to 8\,Hz by requiring the above coincidence.

\section{Calibration}

In order to determine the sensitivity of the telescope, the efficiency of
the light collecting system is the most important parameter to be precisely
measured.
Electromagnetic showers can be accurately simulated by computer code
and cosmic ray showers can be observed in OFF-source observations 
and generated using Monte-Carlo codes such as GEANT.
We measured the point spread function (PSF) 
and reflectivity of each  mirror segment
during their production and  the quantum efficiency of
each photomultiplier (PMT) pixel was also measured.
Some deterioration in these values after the construction of the
telescope is inevitable, and was monitored 
by regular PSF measurements of bright stars. 
Light collection efficiencies were monitored via optical systems
such as photo-diodes and commercial CCD cameras observing stars
both directly, and after reflection by the mirrors.
These showed gradual degradation in the form of broadening spot
sizes and decreasing reflectivities.
However, these data only relate to the reflectors, and we need to know
the total performance of the whole system including the cameras.

In order to measure these, we need ``standard candles''; sources of
Cherenkov photons or non-variable gamma-ray sources such as the Crab Nebula.
Here, we introduce our calibration procedure using these methods.

\subsection{Muon-ring Analysis}

The ideal light source is one that produces the same Cherenkov radiation
that the telescope is designed to detect.
Sub-TeV cosmic rays produce hadronic showers in the upper atmosphere at a
depth of around three inelastic interaction lengths, corresponding
to an altitude of $\sim$10\,km.
Secondary hadrons are produced and interact further, 
with only gamma-rays, electrons and muons surviving to near ground level.
Of these, relativistic muons radiate Cherenkov rings near the Earth's 
surface, and these produce distinct complete or partial ring 
images on the camera.

\begin{figure}
\plotone{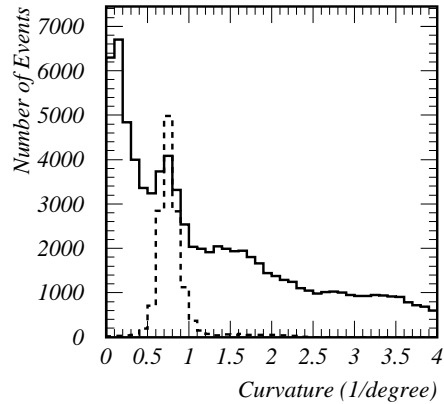}
\caption{Curvature distribution for the circular fitting results of all events
for T3. The solid histogram is for all events and the dashed one for
the selected events (with the vertical scale multiplied by 5).
}
\label{curvature_all_t3}
\end{figure}

\begin{figure}
\plotone{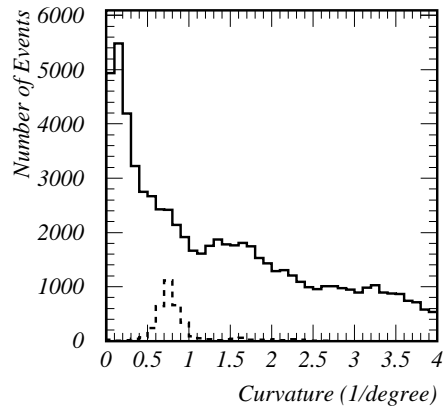}
\caption{Curvature distribution for the circular fitting results on all events
for T2. The solid histogram is for all events and the dashed one for 
the selected events (with the vertical scale multiplied by 5).
}
\label{curvature_all_t2}
\end{figure}

The contribution of these rings to the trigger rate of our telescopes
is critical,
because small Cherenkov images resemble 
gamma-ray images.
Local muon ring images can, however, be removed by requiring the 
coincidence of two or more telescopes separated by $\sim$100\,m.
Here, for calibration, we select images produced by
muons at altitudes of 100--200\,m.
This can be done by selecting a somewhat larger ``arc-length".
Arc-length is measured after fitting a circle to the images, and
is the fitted radius multiplied by the recorded opening
angle of the image.
If larger images are required, we can simply restrict the selection
to geometrically closer images such
as those with altitudes $<$100--200\,m.
The inverse of the radius (curvature) should have a
Gaussian distribution, so we therefore plot the
curvature as a solid histogram in Fig.~\ref{curvature_all_t3}.
A peak can be seen around 1/1.3 [1/degree] which corresponds to the
inverse of the Cherenkov angle initiated by relativistic muons at
normal temperature and pressure.
The distribution for T2 (which was constructed before T3), however, 
was not as good, indicating some degradation in its performance,
as shown by the solid histogram in Fig.~\ref{curvature_all_t2}.
These data were taken in 2003 December. We need to know our light
collection efficiency in each observing period, i.e., we
need to improve the S/N ratio to select a pure muon-ring sample only.
The following are the cut criteria used to achieve this.
\begin{itemize}
\item Hit threshold more than 1 photo-electron (p.e.).
\item TDC range within $\pm$30\,ns from the mean arrival time of the event.
\item At least 15 triggered (or ``hit'') PMTs, with a clustering 
requirement that there be at most two non-triggered PMTs between
nearest-neighbor hit PMTs.
\item A circle, or ring segment, can be fitted to the image.
\item The circle has an ``arc-length" of more than 2 degrees.
\item A $\chi^2$ per hit-pixel, normalized to the pixel size, of less than 1.5.
\end{itemize}
The curvature distributions of the samples meeting these criteria are shown
as dashed histograms
in Figs.~\ref{curvature_all_t3} and \ref{curvature_all_t2}.
Muon ring events are selected with a good S/N ratio.
Geometrically, the arc-length cut restricts selection to events 
occurring at $<$100--200\,m. 
The statistics of the accepted events are sufficient for  
calibration to be undertaken with only several hours data.

There is some dependence of light yield on the atmospheric
temperature; however, this is monitored and recorded every second.
The night-time temperature near Woomera ranges from near 0\,C
in winter to over 30\,C in summer. 
The systematic error,
which includes the reflectivity of 
the present mirror, the PSF, and the quantum efficiency of the PMTs,
is thought to be at the 5\% level of the 
total light collection efficiency.
The light collection efficiencies can be derived from the ``size/arc-length"
distribution
and the PSFs by the $\chi^2$ of the ring fits.
The observation periods for the Crab Nebula and Vela Pulsar data
considered here were
between the end of 2003 and the beginning of 2004.

\begin{figure}
\epsscale{0.7}
\plotone{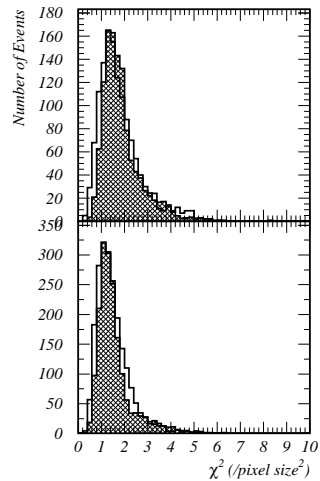}
\caption{
Distributions of $\chi^2$ for the experimental data and the tuned Monte-Carlo.
The solid histograms are the experimental data and the hatched the Monte-Carlo.
The upper panel is for T2 and the lower T3.
}
\label{f_mu_spot}
\end{figure}

\begin{figure}
\plotone{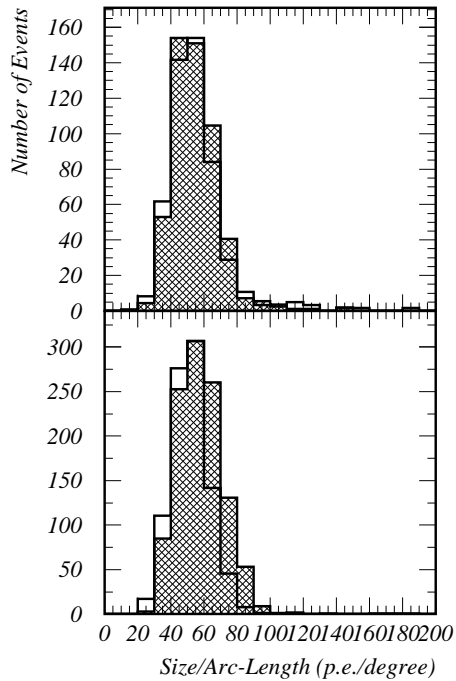}
\caption{
Distributions of size/arc-length 
for the experimental data and the tuned Monte-Carlo.
The solid histograms are the experiment and the hatched the Monte-Carlo.
The upper panel is for T2 and the lower T3.
}
\label{f_mu_size}
\end{figure}

%
The calibrated light collection efficiencies for T2 and T3 
were both 70\%, with 5\% systematic errors.
These values are the ratios of the efficiencies derived during the observation
period and those measured at the production time, 
i.e., the deterioration factors.
The PSFs were 
approximated by Gaussian distributions with standard deviations of
0.14, and 0.12 degrees, respectively, 
which are somewhat 
larger than the results obtained via bright star measurements at the
initial installation time.
(The corresponding values for T4, which was just coming on-line,
were 85\% and 0.09 degrees, respectively.)
The $\chi^2$ distributions for the experiments (the solid
histogram) and the tuned Monte-Carlo simulations (the hatched area)
are shown in Fig.~\ref{f_mu_spot} for T2 (upper panel) and T3 (lower panel).
The PSF of T2 is worse than that of T3.
The size/arc-length distributions for the experiments (the solid
histogram) and the tuned Monte-Carlo (the hatched area)
are shown in Fig.~\ref{f_mu_size} for T2 (upper panel) and T3 (lower panel).
The PSFs obtained for the three telescopes are somewhat
larger than those of H.E.S.S., 
corresponding to a relatively lower
cosmic-ray rejection efficiency. The resulting S/N ratio for
gamma-rays is discussed in the next section;
however, it is sufficient to detect gamma-ray sources 
with flux levels flux of $\sim$0.1~Crab in several tens of hours.

The measured efficiencies do not depend strongly on
climate or atmospheric conditions, as they are local (near-surface) 
phenomena.
Thus uncertainty due to Mie scattering, for instance, still remains.
This is thought to be significant at the 10\% level; however, 
the average effect can be gauged in the Crab Nebula data
described in the following section.
In fact, data taken during cloudy nights showed a better S/N ratio
in the curvature distribution.
As the muon rings are a local phenomenon, 
there is naturally no dependence on the 
elevation angle of the observation, whereas the
cosmic ray rate varies strongly with elevation.
Using these characteristics, we were able to measure the
degradation as a function of time for the 
three telescopes (T2, T3, and T4), with the light collection efficiency 
being found to decrease by 5\% per year.
Hereafter in the analyses of the individual sources, we used 
muon-ring data from the
corresponding periods to tune our Monte-Carlo code.

\subsection{Crab Analysis}

\begin{figure}
\plotone{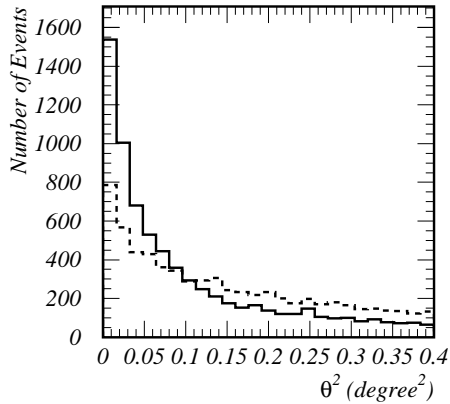}
\caption{
$\theta^2$ distributions from Monte-Carlo simulations. 
The dashed line is obtained without
the fitting procedure described in the text and the solid line is that
with it.
}
\label{f_fit}
\end{figure}

\begin{figure}
\plotone{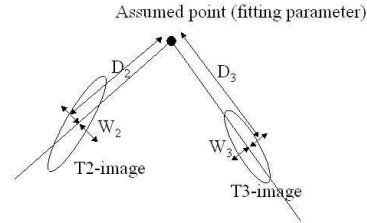}
\caption{
Definitions of parameter $W_T$ and $D_T$ which are used
in the fitting procedure described in the text.
}
\label{ipfit}
\end{figure}

The description on the Crab data used in the following section can be found
in Appendix \ref{a_crab_obs}.
At first, the Hillas moments were calculated for each telescope's image.
The intersection of the major axes of the two images is the incident direction
of the gamma-ray in the stereoscopic observation.
The $\theta^2$ distribution of the Monte-Carlo data for this observation
condition
calculated from the intersection points
is plotted in Fig.~\ref{f_fit} (the dashed line).
The angular resolution is, as expected,
several times worse than that of H.E.S.S.\ \citep{hess_gc},
due to the larger PSFs described above.
The Crab Nebula observations were carried out at relatively
low elevation angles
and the opening angle between the T2 and T3 images are typically small,
resulting in many parallel images.
This results in an increased uncertainty in the intersection point
of the two images.
To avoid this, we 
reanalyzed the data
with a constraint
on the distances between images' center of gravity and 
the intersection point.
(H.E.S.S.\ observations are made at elevation angles
typically 10 degrees higher and with smaller mirror PSF,
resulting a good S/N ratio without this constraint.
If the PSFs of mirrors were as small as specified
in the original design, we could also avoid this procedure.)
The $\chi^2$ was defined as
$$\sum_{T=2,3}\left[W_T^2+\frac{(D_T-\langle Distance\rangle)^2}
{\sigma_{Distance}^2}\right],$$
where $W_T$ is the {\em width} seen from the intersection point, 
and $D_T$ is the {\em distance}
between the image center of gravity and the intersection for each telescope
(Fig.~\ref{ipfit}), 
$\langle Distance\rangle$ is the mean {\em distance} obtained by
Monte-Carlo simulations for gamma-rays, 
and $\sigma_{Distance}$ is its standard deviation.
The improved $\theta^2$ distribution is shown as the solid line
in Fig.~\ref{f_fit}.
The number of events with $\theta^2<$\,0.05\,degree$^2$ increases 
by a factor of 1.8.
Note that the angular resolution was estimated to be 0.23 degree 
and 0.23$^2$ is roughly 0.05\,degree$^2$.

\begin{figure}
\plotone{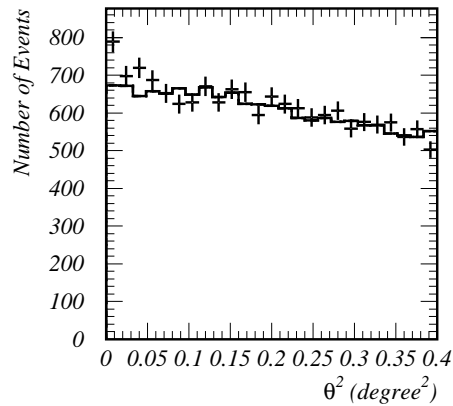}
\caption{
$\theta^2$ distribution for the standard analysis of Crab Nebula data.
The points with error bars are from the ON-source region and the
solid histogram is the background region.
}
\label{f_sq}
\end{figure}

We first used the conventional (``square cuts") method as follows;
we accepted events with 
{\em width}\,$<$\,0.2 (for T2)
and 0.15 (for T3), and {\em length}\,$<$\,0.3 (for T2) and 0.25 (for T3),
respectively, where all values are in units of degrees. The 
cut values differ for T2 and T3 due to the PSF
differences.
As the $\theta$ resolution is 0.23 degree, we 
took six background points on the 0.5-degree-radius circle 
from the pointing direction.
The resulting $\theta^2$ distributions for the ON-source points
(the points with error bars)
and the OFF-source points (the solid histogram)
are shown in Fig.~\ref{f_sq}.
The background normalization factor is 1/6, equal to the inverse
of the number
of background points in these wobble mode observations.
The number of excess events with $\theta^2<0.05$ degree$^2$ is
(2207$-$1990=) 216, where the numbers in parentheses are
the on-source and estimated background counts, respectively.
This corresponds to a \cite{lima} significance of 4.4$\sigma$.
The number of events predicted by our Monte-Carlo simulation,
using the \cite{hegra_crab} flux, an $E^{-2.5}$ gamma-ray energy spectrum,
and minimum and maximum 
gamma-ray energies of 500\,GeV and 20\,TeV,
was 195 events.
(For comparison, we also undertook ``mono'', i.e., single telescope, analyses
using the regular CANGAROO-II procedures. 
Excesses for both telescopes were found using
both by square cut and Likelihood cut analyses.
The statistical significances of these excesses were 
at the 3$\sigma$ level,
confirming the power of the stereo technique.)

\begin{figure}
\plotone{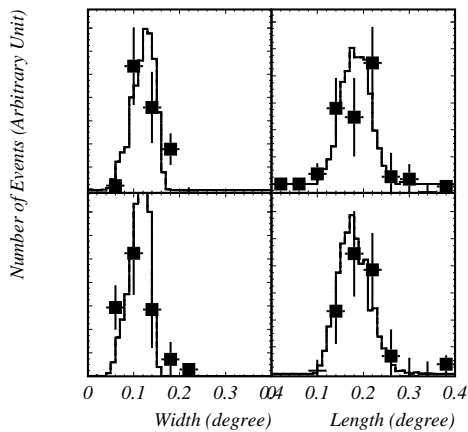}
\caption {
Hillas moment distributions: upper-left (T2 {\em widths}), upper-right
(T2 {\em lengths}), lower-left (T3 {\em widths}), 
and lower-right (T3 {\em lengths}). 
The points with error bars were obtained after the subtraction
of background events. The solid histograms are the Monte-Carlo
predictions where the total number of entries were normalized to those
of the observations.
}
\label{f_shape}
\end{figure}

Using the events with $\theta^2 < 0.05$ degree$^2$, we can
derive the Hillas moment distributions after background
subtractions.
This provides a check of how well the Monte-Carlo simulations agree
with the real gamma-ray data.
These are shown in Fig.~\ref{f_shape}.
Our Monte-Carlo simulations are consistent with the data
within statistical uncertainty.
Note that the {\em width} distribution is sensitive to the PSF
of the mirror system and that that of T3 is better than T2.

\begin{figure}
\plotone{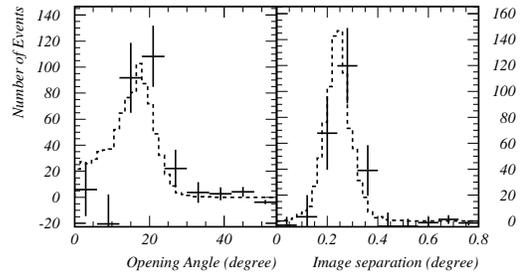}
\caption{
Distributions of opening angle and separation angle of two images.
These were obtained also via the subtraction described above for
the Crab data.
}
\label{f_shape2}
\end{figure}

\begin{figure}
\plotone{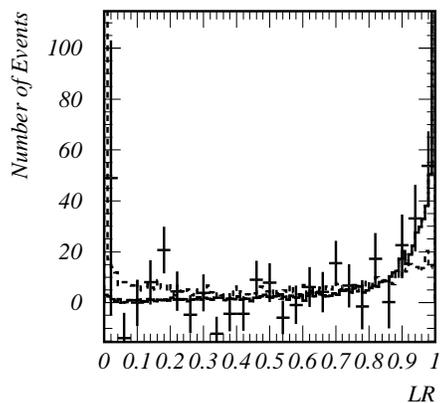}
\caption{
Likelihood Ratio ($LR$) distributions. The points with error bars
are from Crab Nebula gamma-rays, the solid histogram from the gamma-ray
Monte-Carlo, and the dashed histogram from all events.
The latter two distributions are 
normalized to the total number of the observed excess.
}
\label{f_l}
\end{figure}

After this standard analysis, we investigated two
bias-free analyses.
The first is the Likelihood method introduced by \cite{enomoto_nature}.
In the standard ``square cuts", there are four
parameters on which cuts can be made, and in the absence of
a strong gamma-ray source or detailed simulations accurately
incorporating the characteristics of the telescope hardware,
some freedom in choosing the exact cuts is available.
\citep[see, e.g., the discussion in][]{dazeley2}.
In the Likelihood method, 
we proceed by making probability density functions (PDFs) from
the histograms of {\em width} and {\em length}, for the two telescopes,
and some other
parameters such as the opening angle and the distance between two images'
center of gravity (image separation) (Fig.~\ref{f_shape2}).
In order to treat the energy dependence of these
parameters, we used two-dimensional histograms:
the PDFs are therefore 2-D functions.
For the gamma-ray sample, we used the data from Monte-Carlo simulations,
and for the cosmic ray background sample, we used real observation
data, since the ``contamination'' by 
gamma-rays is much less than 1\%.
Thus for each event, we can derive a probability ($L$) for the
event being initiated by a gamma-ray or a cosmic ray,
where normalization ambiguities still remain.
The Likelihood ratio $LR$ is defined as
$$LR=\frac{L(gamma-ray)}{L(background)+L(gamma-ray)}.$$
The distributions for the Crab Nebula data are shown in Fig.~\ref{f_l}.
The observational data are in reasonable agreement with the Monte-Carlo
predictions.
Observationally, the optimal cut was found to be $LR>$0.9.
The number of excess events is (390$-$289=) 101
(a Li and Ma significance of 5.2\,$\sigma$)
where the Monte-Carlo expectation
is 121 events.
The determination of the cut position is, however, still somewhat subjective.

\begin{figure}
\plotone{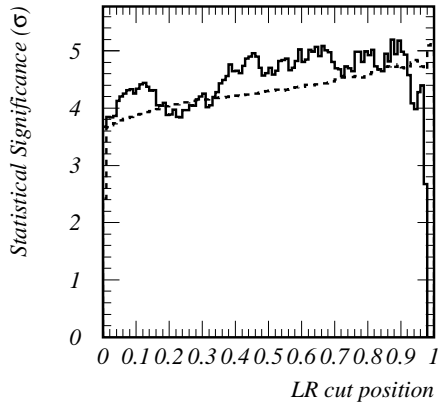}
\caption{
Statistical significance of the Crab excess versus $LR$ cut position
(the solid line). The dashed curve is the figure of merit
calculated as described in the text. The normalization
is arbitrary.
}
\label{f_foml}
\end{figure}

\begin{figure}
\plotone{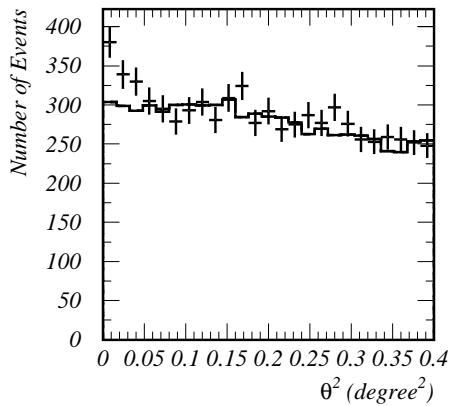}
\caption{
$\theta^2$ distribution for Crab Nebula data 
obtained using the $LR>$ 0.5 cut.
}
\label{f_l05}
\end{figure}

A simple figure of merit ---
the number of the accepted events in the Monte-Carlo simulation
divided by the square-root of those in the observation --- is 
plotted as a function of $LR$ in Fig.~\ref{f_foml}.
The Monte-Carlo prediction is a smoothly increasing function
of the cut position (the dashed line).
On the other hand, the statistical
significance of the observed excess is shown by the solid line.
The two show good agreement over a wide range; however, some discrepancy
is apparent in the extremely tight cut positions.
For example, a $LR$ cut at 0.5 gives an excess (Fig.~\ref{f_l05}),
of (1049$-$895=) 154 events (4.6$\sigma$) where the expectation is 182 events.
Further fine-tuning of the Monte-Carlo code is
necessary in the future and such tight cuts will for now
be avoided.
The optimal cut position of 0.9 is located on the edge of this tight cut
region.

\begin{figure}
\plotone{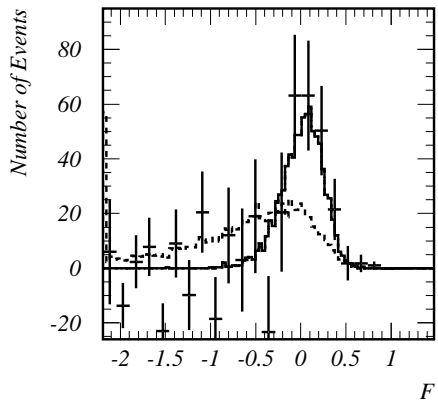}
\caption{
Distributions of the Fisher Discriminant, $F$.
The points with error bars are obtained from the
observation, the solid histogram from the gamma-ray
Monte-Carlo, and the dashed from all events representing
background behavior.
}
\label{f_f}
\end{figure}

We now investigate an alternative approach for the comparison of
observational and Monte-Carlo data: the Fisher Discriminant \citep{fisher}.
When we use multi-parameters:
$$\vec{P}=(Vector~of~Hillas~moments~for~two~telescopes)$$
and assume that a linear combination of
$$F=\vec{\alpha}\cdot\vec{P},$$
provides the
best separation between signal and background, then the set of
linear coefficients ($\vec{\alpha}$) should be uniquely determined as
$$\vec{\alpha}=\frac{\vec{\mu}_{sig}-\vec{\mu}_{BG}}{E_{sig}+E_{BG}},$$
where $\vec{\mu}$ is a vector of the mean value of $\vec{P}$ for each
sample, i.e., $=\langle\vec{P}\rangle$, and $E$ is their error matrix,
i.e., $=\langle\vec{P}\vec{P}^T\rangle-\langle\vec{P}\rangle\langle\vec{P}^T\rangle$.
The values of
$\vec{\mu}_{sig}$, $\vec{\mu}_{BG}$,
$E_{sig}$, and $E_{BG}$ can be calculated from the
Monte-Carlo gamma-ray events for signal and observational data for background.
Our purpose is to separate ``sharp images'' from ``smeared ones".
This method is regularly used in high-energy physics experiments, such
as the $B$-factory's ``Super Fox-Wolfram moment" \citep{two}, 
in order to separate spherical events from
jet-like events.
Here, the assumption strongly depends on which linear combination
is best. We must, therefore, select parameters which are
similar to each other. {\em Width} and {\em length} are
both second order cumulative moments of shower images
and thus a linear combination is a reasonable assumption.
We used four image parameters: {\em T2-width}, {\em T3-width},
{\em T2-length}, and {\em T3-length}.
Here, the energy dependence of {\em width} and {\em length} for each telescope
were corrected using the energy estimated from the summation of 
ADC values of hit pixels for that telescope
using the Monte-Carlo expectations. This correction function was
a second-order polynomial and we carry out an offset correction
using it, i.e.,
$$\vec{P}=\vec{P}_{raw}-\vec{a}_0-\vec{a}_1log(\sum ADC)-
\vec{a}_2log(\sum ADC)^2,$$ 
where $\vec{a_i}$s were determined from the two-dimensional
plots of raw ``Hillas moments" versus $log(\sum ADC)$ using the 
Monte-Carlo gamma-ray simulations and $\vec{P}_{raw}$ is a vector made
of the raw values of ``Hillas moments", respectively.
The corrected
{\em width} and {\em length} are distributed around zero and
the $\langle\vec{P}_{sig}\rangle$ is automatically located exactly
at zero, 
$$\langle\vec{P}_{sig}\rangle=0$$
resulting in the Monte-Carlo expectation of $F$
location being zero:
$$\langle F_{sig} \rangle = 0.$$
This removes cut-selection bias: a cut at zero ensures 50\%
acceptance (for gamma-rays) automatically, a typical 
value adopted in IACT analyses. From the $\theta^2$ 
Crab Nebula excess we obtain the following
distribution of Fisher Discriminant, $F$, as shown in Fig.~\ref{f_f}.
The behavior of the distributions for gamma-ray signals, 
background events, and Monte Carlo expectations agree reasonably
well and are similar or a little bit better 
than the result obtained for the Likelihood method.
This figure should be compared with Fig.~2 in the recently published
H.E.S.S.\ paper for the distribution of their 
{\em mean reduced scaled width}  \citep{hess_1303}.
The H.E.S.S.\ discrimination is better; however, improved results can be
expected for CANGAROO-III if the quality of mirrors were improved to the
levels of H.E.S.S.\ or MAGIC.

\begin{figure}
\plotone{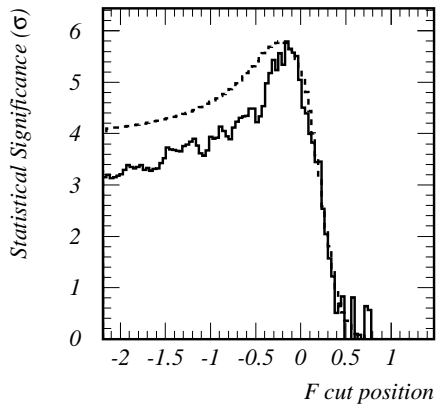}
\caption{
Statistical significance of the Crab Nebula excess versus $F$ cut position
(the solid line). The dashed curve is the figure of merit
calculated as described in the text. The normalization
is arbitrary.
}
\label{f_fomf}
\end{figure}

\begin{figure}
\plotone{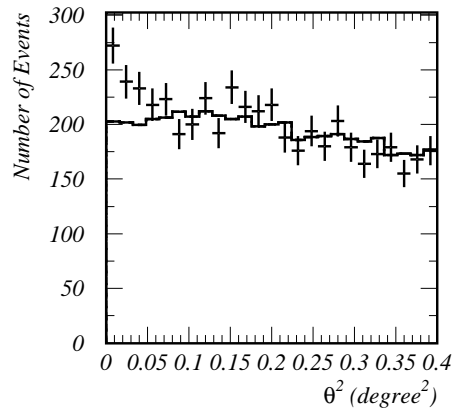}
\caption{
The final $\theta^2$ plot for the Crab Nebula, using the Fisher Discriminant 
analysis with the cut at zero.
}
\label{f_zerof}
\end{figure}

We consider the figure of merit for this method in Fig.~\ref{f_fomf}.
This time, the agreement is very good in all regions.
Although the cut at $F>-$0.16 showed the best statistical
significance, 5.8$\sigma$, with an 
excess of (1177$-$974=) 203 events, 
where the Monte-Carlo expectation is 162 events.
A conservative choice of cut point is zero which,
as noted above,
is exactly the mean expected position for a gamma-ray
acceptance of 50\%.
In the following we use the Fisher Discriminant analysis as the default,
with a cut position at zero,
thus removing all subjective or potentially biased 
elements from the analysis.
The excess is (744$-$604=) 140 events, where the expectation is 110 events.
The final $\theta^2$ distribution is shown in Fig.~\ref{f_zerof}.

\begin{figure}
\plotone{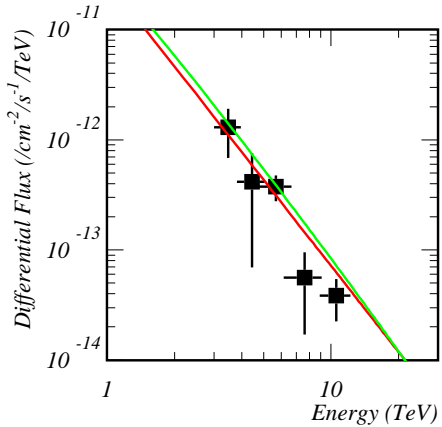}
\caption{
Differential gamma-ray flux from the Crab Nebula as a function of energy.
The red line is the HEGRA result \citep{hegra_crab} 
and the green is the Whipple result \citep{whipple_crab}.
}
\label{f_dfcrab}
\end{figure}

In this analysis, we estimated the gamma-ray energy
using the average of the T2 and T3 summations of ADC values,
which probably provides the optimal energy resolution 
for this analysis.
The usage of the intersection constraint, 
due to finite point spread function of the individual mirror
segments, does not
allow a core-distance correction of energy.
The overall energy resolution was considered to be 30\%
from the simulation study.
We show the differential flux for the Crab Nebula
in Fig.~\ref{f_dfcrab}.
The points with error bars are our data where the errors
are statistical. The systematic error is considered
to be 15\% at this stage.
HEGRA \citep{hegra_crab}  and Whipple \citep{whipple_crab}
results are also shown.
The results agree within the statistical and systematic errors.
The deviation in the CANGAROO flux at higher energies may be due to 
saturation effects, which are not yet fully implemented in our
Monte-Carlo code; however, this in not an important
consideration for Vela observations, which are the main focus
of this paper.
We leave further consideration of this for the future, noting
that while the analysis template defined here 
for CANGAROO-III is, at present, most suitable, it  
may be able to be further optimized with refined Monte-Carlo simulations.

\begin{figure}
\plotone{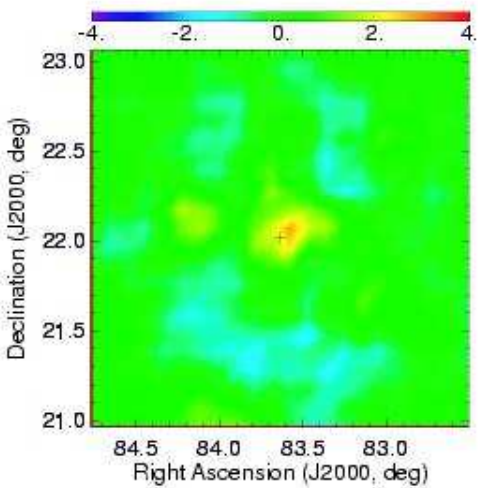}
\caption{
Excess number plot in the equatorial frame.
The unit is events per $0.1\times 0.1$ degree$^2$ in the FOV coordinates.
The cross indicates the center of the Crab Nebula.
The resolution is considered to be 0.23 degrees
in declination in this frame.
}
\label{f_profile}
\end{figure}

We show the 2D profile of gamma-ray images in Fig.~\ref{f_profile}.
The angular resolution 
is estimated to be 0.23 degrees (1$\sigma$).
Here, only high energy events ($E>$10.6\,TeV) 
are plotted in order to improve the
S/N ratio and to remove uncertainties in the background subtraction.
The background level at each position was estimated using
the events with values of $F$ below $\langle F_{BG} \rangle$
which are considered to be mostly background protons.
Here, however, we need to note that the subtraction template for the background
was made using events with low $F$ sample. 
The Fisher Discriminant uses a linear combination of
the energy-corrected {\em width} and {\em length}
for T2 and T3, respectively.
The subtraction sample for morphological studies may tend to be overwhelmed
by the larger sized events which have poorer angular resolutions.
The important thing is that the background sample tends to be flatter than
the gamma-ray events. Therefore, we only show a very restricted area in this
figure. An improved analysis, required for diffuse gamma-ray detection,
which will be introduced later.


As a result of the considerations above,
the analysis template is:
\begin{itemize}
\item the Fisher Discriminant is adopted.
\item the cut position on $F$ is exactly at zero.
\item others parameters,
such as elevation cut and shower-rate cut, 
are determined as appropriate for the source under
investigation (depending on its declination and galactic coordinates).
\end{itemize}

\section{Observations of the Vela Pulsar}

The Vela Pulsar was observed between 2004 January 17 and February 25.
In total, the preselected data correspond to an analyzable
period of 1311 min.\ after the elevation and cloud cuts,
where the minimum elevation angle was set at 60~degrees
and the shower rate at 9\,Hz.
The mean elevation angle was 70.9~degrees, corresponding
to an energy threshold of 600\,GeV.
The observations were carried out using the same wobble mode
as for the Crab Nebula observations.
In this period, T2 and T3 were in operation and 
we analyzed the stereo data from these two telescopes.
As the Vela Pulsar is at a declination of $-$45$^\circ$, the
relative orientation of the two telescopes does not present
any problems.

\begin{figure}
\plotone{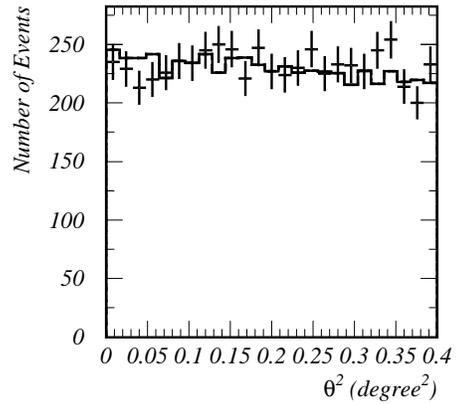}
\caption{
The $\theta^2$ distribution for Vela Pulsar observations
resulting from application of the analysis template
described in \S3.3.
}
\label{f_fzero}
\end{figure}

\begin{figure}
\plotone{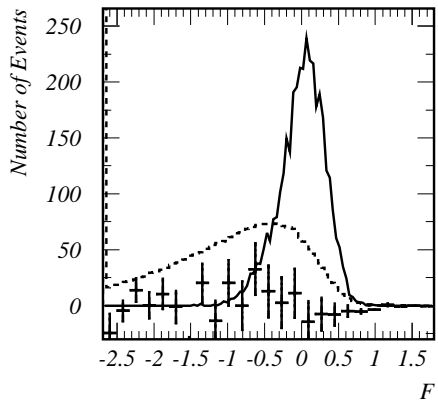}
\caption{
$F$ distributions.
The points with error bars are obtained from the
observations, the solid curve from the gamma-ray
Monte-Carlo simulations, and the dashed histogram
from all observed events to represent the cosmic ray
background behavior.
The Monte-Carlo histogram is normalized to a 100\%\,Crab Nebula flux level.
}
\label{f_f_vela}
\end{figure}

We used the optimized analysis procedure described in the
previous section and so there are no {\em a posteriori} trials
to consider in the interpretation of results --- it is
a ``blind analysis".
The resulting $\theta^2$ distribution is shown in Fig.~\ref{f_fzero}.
After the background subtraction, we obtained an excess
of (677$-$722=) $-$45$\pm$29 events.
Monte Carlo simulations, with minimum and maximum gamma-ray energies
of 100\,GeV and 20\,TeV, respectively, and an energy spectrum
proportional to $E^{-2.5}$, predict 394 events would 
be detected for a 100\%\,Crab level gamma-ray source.
The CANGAROO-I flux was 60\% of the Crab Nebula flux.
The $F$ distribution, obtained using subtraction,
is shown in Fig.~\ref{f_f_vela}.

There is no excess of gamma-ray--like events around the predicted
region, offset by 0.13~degrees from the Vela Pulsar. 
Thus observations with significantly improved instrumentation
and a robust analysis procedure do not support the previous
claim for TeV gamma-ray emission from this region.

\begin{figure}
\plotone{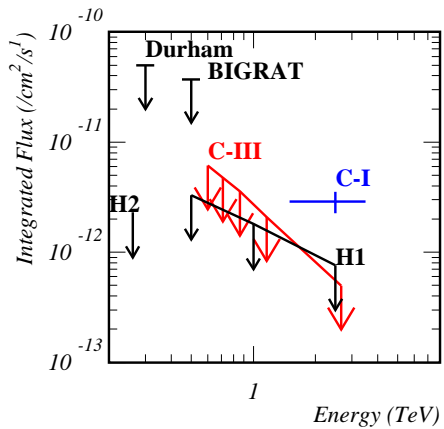}
\caption{
2$\sigma$ upper limits for the gamma-ray flux.
C-I represents the CANGAROO-I result \citep{yoshikoshi}
and C-III indicates the CANGAROO-III result reported here.
Also shown are Durham \citep{chadwick}, 
BIGRAT \citep{dazeley3} and H.E.S.S.\ preliminary
results \citep{gamma2004} \citep{icrc2005}.
}
\label{f_fvela}
\end{figure}

The 2$\sigma$ upper limits are shown in Fig.~\ref{f_fvela},
together with the results of other observations.
The upper limits in the figure are a factor of 5 below the
CANGAROO-I result. This analysis has used the point
offset by 0.13 degrees to the south-east of the Vela Pulsar, i.e.,
offset is ($\Delta$RA,$\Delta$dec)=(0.14$^\circ$,$-$0.1$^\circ$)
from the pulsar position.
This position 
which was the maximum of the excess detected with the CANGAROO-I
telescope: an analysis at the position of the Vela Pulsar position 
yields similar upper limits.
These are summarized in Table \ref{t_flux}.
\begin{table}
\begin{center}
\caption{
2-$\sigma$ upper limits to the integrated fluxes.
}
\label{t_flux}
\begin{tabular}{c c c}
\tableline\tableline
             & $F$($>E$) at        & $F$($>E$) at  \\
Energy, $E$  & offset position     &  pulsar position \\
(GeV)        & (cm$^{-2}$s$^{-1}$) & (cm$^{-2}$s$^{-1}$) \\
\tableline
~600 & ~ ~ $<5.8\times 10^{-12}$ ~ ~ & ~ ~ $<5.8\times 10^{-12}$ ~ ~ \\
~710 & ~ ~ $<4.5\times 10^{-12}$ ~ ~ & ~ ~ $<4.5\times 10^{-12}$ ~ ~ \\
~860 & ~ ~ $<3.4\times 10^{-12}$ ~ ~ & ~ ~ $<3.4\times 10^{-12}$ ~ ~ \\
1200 & ~ ~ $<2.0\times 10^{-12}$ ~ ~ & ~ ~ $<2.4\times 10^{-12}$ ~ ~ \\
2700 & ~ ~ $<4.7\times 10^{-13}$ ~ ~ & ~ ~ $<4.7\times 10^{-13}$ ~ ~ \\
\tableline\tableline
\end{tabular}
\end{center}
\end{table}

\section{Vela Pulsar Wind Nebula}

Recently, extended TeV gamma-ray emission coincident with the 
Vela Pulsar Wind Nebula (PWN) was reported by the H.E.S.S.\ group 
\citep{icrc2005}. Their preliminary report
claimed that the center of the emission is 
(RA, dec) = (8$^h$35$^m$, $-$45$^\circ$36$'$) 
and that the flux within a 0.6 degree 
radius from this position is 50\% of the Crab Nebula at 1\,TeV. 
They also noted that no gamma-rays were detected 
from the Vela Pulsar, and placed a tight upper limit 
on the pulsar flux at 0.26\,TeV
\citep{gamma2004,icrc2005}.

\begin{figure}
\plotone{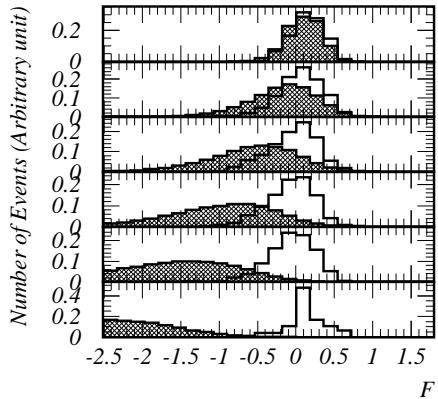}
\caption{
$F$ distributions in various energy regions. The blank histograms are
for Monte Carlo simulations of gamma-rays and the hatched histograms
for background regions of real data. From the top panel to the bottom, 
the central gamma-ray energies are 
540, 780, 1200, 2700, 3300, and 8800\,GeV, respectively.
The histogram entries are normalized to one event in the shown regions.
}
\label{fvse}
\end{figure}

Thus far, we have focused on a point source analysis based on the 
wobble mode observations. 
The peak PWN source position coincides with the one of two wobble
pointing direction in the coordinate of the field of view of the camera.
We, therefore, can not carry out background estimation using the usual
wobble method.
In this section we undertake an optimized analysis for an extended source.
Another difficulty is that we don't have sufficient statistics 
for OFF source regions, and so
the background subtractions should be carried out using the ON source
data runs. Gamma-ray--like events can be extracted by fitting 
position-by-position $F$
distributions under the assumption that
gamma-rays obey the Monte-Carlo predictions, the proton background
follows the average $F$ distribution of all directions, and the total
distribution is a linear combination of those two.
The separation between those two distributions is likely to be worse
at lower energies due to the smaller image sizes.
These distributions are plotted in Figs~\ref{fvse} for 
various energy ranges. 
From the upper to lower panels, 
the central energies of the energy bins are
540, 780, 1200, 2700, 3300, and 8800\,GeV, respectively.
The blank histograms are gamma-rays and the hatched, protons.
As shown in figure, the separation begins at 780\,GeV and
becomes significant at 1200\,GeV. Therefore, we first analyzed events with
energy greater than 1200\,GeV.

\begin{figure}
\plotone{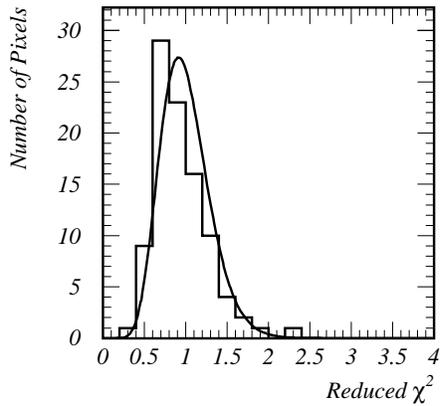}
\caption{
Reduced $\chi^2$ distribution for various arrival directions (the histogram).
The pixel size for each direction is 0.2\,$\times$\,0.2\,degree$^2$. 
The curve is the expected distribution for the number of
degrees of freedom, 23.
}
\label{fchisq}
\end{figure}

Then we checked the directional dependence of the $F$ distributions.
The field of view was segmented into 0.2$\times$0.2~degree$^2$ regions
and each $F$ distribution was compared with the average.
The reduced $\chi^2$ distribution is shown by the histogram in
Fig.~\ref{fchisq}. 
The curve shown is the predicted distribution
for the 23 degrees of freedom. There is good agreement,
i.e., there is no significant directional dependence.

\begin{figure}
\plotone{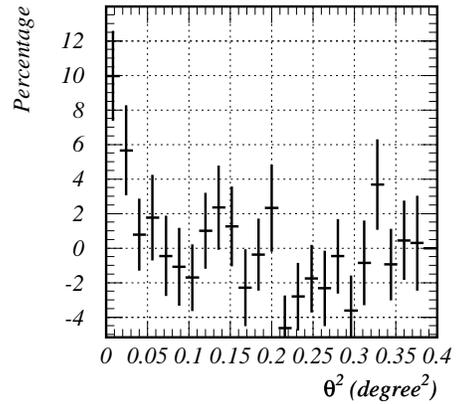}
\caption{
The $\theta^2$ plot obtained by the fitting method
described in the text for Crab Nebula data.
The vertical scale is the percentage of gamma-ray--like events relative to 
all events.
}
\label{crab}
\end{figure}

This method was then checked with Crab Nebula data. The $F$ distributions
at various $\theta^2$ slices were taken at energies greater
than 5.7\,TeV. The background $F$ distribution was obtained in the
higher $\theta^2$ region, 0.1--0.3~degree$^2$.
The only fitting parameter is the percentage of gamma-ray--like events
relative to the total events.
The result is shown in Fig.~\ref{crab}.
The statistical significance of the peak is 4.0$\sigma$, while the
ordinary wobble analysis gave a 3.6$\sigma$ excess.

\begin{figure}
\plotone{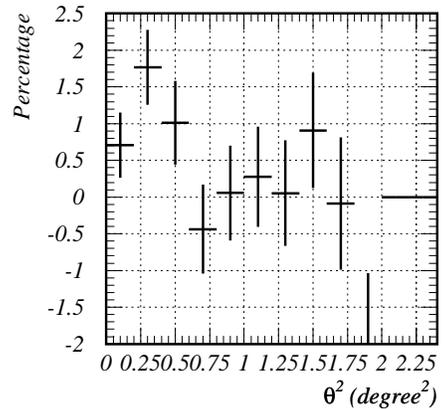}
\caption{
The wide range $\theta^2$ plot for the Vela PWN region, where
$\theta$ was calculated from (RA, dec) = (8$^h$35$^m$, $-$45$^\circ$36'),
i.e., the peak of the emission detected by H.E.S.S. 
The gamma-ray--like events were
extracted by the fitting procedure described in the text.
}
\label{f2_off}
\end{figure}

Having demonstrated the validity of this method,
we carried out an analysis of the Vela PWN region.
The H.E.S.S.\ group detected a gamma-ray excess extended over 
a 0.6~degree radius from 
the center of the emission
[(RA, dec) = (8$^h$35$^m$, $-$45$^\circ$36$'$)]. 
In our case, the angular resolution (0.23~degree) 
is significantly worse than
that of H.E.S.S. We, therefore, chose the background region
to be more than 0.8 degree from the center.
The result of fitting is shown in Fig.~\ref{f2_off}. 
An excess was observed
at $\theta^2 < 0.6$ degree$^2$ around the center of Vela PWN region. 
The excess radius is marginally consistent
with H.E.S.S.\ considering our angular resolution. 
The total number of gamma-ray--like events is 561$\pm$114.

\begin{figure}
\plotone{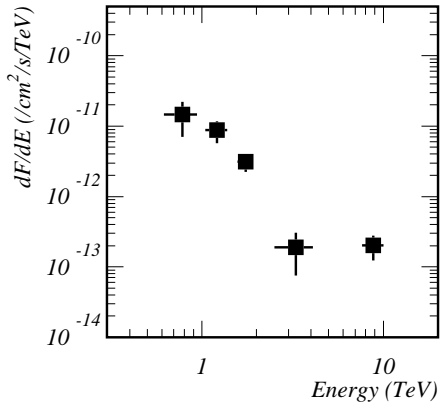}
\caption{
Gamma-ray flux in the Vela PWN region.
}
\label{fitdflux_off}
\end{figure}

The differential flux was obtained and is shown in Fig.~\ref{fitdflux_off}.
Although the statistics are poor,
the spectrum looks hard, consistent with the preliminary H.E.S.S.\ results,
and the fluxes are in general agreement.
This excess is below the 5$\sigma$ level 
generally required for a firm detection \citep{weekes} and is
spatially 
extended over a significant
portion of our field of view,
and thus a high resolution (0.1 degree) morphology map
is not justified.
We are, however, able to offer supporting evidence of the 
H.E.S.S.\ result and, in this light, are planning to observe
the PWN region in more detail next year.


\section{Conclusion}

We have observed the Vela Pulsar region from 2004 January 17 to February 25
with the CANGAROO-III stereoscopic imaging Cherenkov telescopes.
At that time, two telescopes (T2 and T3) were in operation
and events coincident to the two telescopes were analyzed. 
Calibration was performed using muon rings and the performance
of the telescopes confirmed with observations of the Crab Nebula.
The estimated energy threshold for this analysis was 600\,GeV.
The use of the Fisher Discriminant has been introduced and 
a template for CANGAROO-III analysis presented.
No significant excess of events was found from the
Vela Pulsar direction or from the peak of the emission detected 
with the CANGAROO-I telescope. 
The upper limits obtained are a factor of five less than 
the CANGAROO-I fluxes (assuming an $E^{-2.5}$ spectrum).
In addition, we have confirmed, at the 4$\sigma$ level
the gamma-ray emission recently reported from the
Vela Pulsar Wind Nebula. The TeV emission from the PWN peaks
$\sim$0.5 degree south of the pulsar position
and has a $\sim$0.6 degree extension, consistent with the H.E.S.S.\ report.
A detailed morphological study of this source will 
require more observations.

\acknowledgments

This work was supported by a Grant-in-Aid for Scientific Research by
the Japan Ministry of Education, Culture, Sports, Science and Technology, 
the Australian Research Council, and by JSPS Research Fellowships.
We thank the Defense Support Center Woomera and BAE Systems.

\appendix
\section{Crab Observation}
\label{a_crab_obs}

Observations of the Crab Nebula were carried out
over the period 2003 December 18 to 28.
The observations were made using the so-called ``wobble mode"
in which the pointing position of each telescope was
alternated between $\pm$0.5~degree in declination from the 
center of the Crab Nebula every 20 minutes \citep{wobble}.
Two telescopes, T2 and T3, were operational at this time.
T3 is located 100\,m to the south-south-east of T2, which is
not ideal for observations of northern sources; however
sufficient data was recorded for a useful analysis to be made.
The trigger rates of the individual telescopes were at most 80\,Hz 
and this was
reduced to 8\,Hz by requiring the above coincidence.
The total observation time was 1122\,min.
Next we required both telescopes to have clusters of PMTs
with five adjacent hits above a 5\,p.e.\ hit threshold,
which reduced the event rate to 6\,Hz.
Looking at the time dependence of this event rate, we can remove data
taken in cloudy conditions. This procedure is exactly the same as 
the ``cloud cut'' used in CANGAROO-II analysis \citep{enomoto_nature}.
Only data taken at elevation angles greater than 30 degrees were used,
resulting in a total of 890 minutes data, with a 
mean elevation angle of 35 degrees.

\end{document}